\documentstyle[aaspp4]{article}

\begin{document}

\slugcomment{Astrophysical Journal Letters, accepted, 23 November 1999}

\title{Detection of Planetary Transits Across a Sun-like Star}
 
\author{David Charbonneau \altaffilmark{1,2},
Timothy M. Brown \altaffilmark{2},
David W. Latham \altaffilmark{1}, and
Michel Mayor \altaffilmark{3}}
\altaffiltext{1}{Harvard-Smithsonian Center for Astrophysics, 
60 Garden Street, Cambridge, MA 02138; dcharbonneau@cfa.harvard.edu, 
dlatham@cfa.harvard.edu}
\altaffiltext{2}{High Altitude Observatory, National Center for 
Atmospheric Research, P. O. Box 3000, Boulder, 
CO 80307-3000; timbrown@hao.ucar.edu.  The National Center for Atmospheric
Research is sponsored by the National Science Foundation.}
\altaffiltext{3}{Observatoire de Gen\`eve, CH-1290 Sauverny,
Switzerland; Michel.Mayor@obs.unige.ch}

\begin{abstract}
We report high precision, high cadence photometric measurements of
the star HD 209458, which is known from radial velocity measurements
to have a planetary mass companion in a close orbit.  We 
detect two separate transit events at times that are consistent
with the radial velocity measurements.  In both cases, the detailed
shape of the transit curve due to both the limb darkening of the
star and the finite size of the planet is clearly evident.  
Assuming stellar parameters of 1.1 $R_{\sun}$ and
1.1 $M_{\sun}$, we find that the data are best interpreted as
a gas giant with a radius of $1.27 \pm 0.02 \ R_{\rm Jup}$ in an orbit
with an inclination of $87.1 \pm 0.2^{\circ}$.  We present
values for the planetary surface gravity, escape velocity, and average
density, and discuss the numerous observations that are warranted
now that a planet is known to transit the disk of its parent star.
\end{abstract}


\section{INTRODUCTION}
Radial velocity surveys of nearby F, G, K and M dwarf stars 
have revealed a class of close-in extrasolar massive planets
that orbit their stars with an orbital separation of 
$a \lesssim 0.1$ AU.  There are currently eleven such
candidates known (Mayor \& Queloz 1995;
Butler et al. 1997; Butler et al. 1998; Fischer et al. 1999;
Mayor et al. 1999; Queloz et al. 1999; Udry et al. 1999; 
Mazeh et al. 2000).  
Prior to the transit results on this star,
the radial velocity method has been the only method by which
we have learned anything about these planets. 
The radial velocity technique measures the period, semi-amplitude, and
eccentricity of the orbit, and by inference the semi-major axis and
minimum mass, dependent upon the assumed value for the stellar mass.
The search to measure the transit photometrically is motivated
by fact that, for a star for which both the radial velocity
and transits are observed, one can estimate both the mass 
(with negligible error due to $\sin i$) and radius of the planet.
These can then be combined to calculate such
critically interesting quantities as the surface gravity and 
average density of the planet, and thus provide the first
constraints on structural models for these low-mass
companions.  Assuming a random alignment of the orbital inclination
to the line-of-sight for a system with $a = 0.05 $ A.U., the chance
of a transiting configuration is roughly 10\%, depending upon the
value of the stellar radius.

In this letter, we present observations of
the photometric dimming of HD 209458, which we attribute to
the passage of the planet across the stellar disk.
Observations of this star covering less than one full transit have
also recently been reported by Henry et al. (1999).
Motivation for observing HD 209458 came from D. W. Latham and M. Mayor
(personal communication) in August 1999.  The star had been observed both
on Keck I with HIRES (Vogt et al. 1994) and with ELODIE on the 1.93-m telescope
at Observatoire de Haute Provence (Baranne et al. 1996) as one of the
targets in two independent searches for extrasolar planets.  When the two
teams discovered that they had both detected low-amplitude velocity
variations, they agreed to pool their efforts.  Additional observations
were then obtained with CORALIE on the new 1.20-m Swiss telescope at La
Silla (Queloz et al. in preparation). The preliminary orbital solution
from the combined data was then used to predict times of transit for the
photometric observations.
A proper interpretation of the transit results requires accurate
estimates of the mass, radius, and limb darkening of HD209458, and of
the observed radial velocity amplitude of the primary.
These issues are addressed in a companion paper (Mazeh et al. 2000,
henceforth M00). 
In what follows, we have used parameters of HD209458 and of its radial
velocity orbit from this source except as noted.

\section{OBSERVATIONS AND DATA REDUCTION}
We undertook a program to observe HD 209458 both to
establish that the star was photometrically stable for the majority
of its orbit (and thus support the hypothesis that the observed 
radial velocity variations were due to an orbiting companion and
not due to some form of stellar variability), and to search for
planetary transits.  
We obtained photometric observations using the STARE Project 
Schmidt camera (focal length = 286 mm,
f/2.9), which images a field 6$^{\circ}$ square onto a 
2034 $\times$ 2034 pixel
CCD with 15 $\mu$ pixels.
This camera was designed to search for planetary transits in large samples
of stars; it is described more fully in Brown \& Kolinski (1999).

We observed HD 209458 for ten nights (UT dates 29, 30 Aug, 
\& 1, 6-9, 11, 13, 16 Sep 1999).  
In order to avoid saturation from the high flux from the star
($V = 7.65$), we defocused the telescope.  The resulting distorted point
spread function caused no problems, as we later analyzed the
images by means of aperture photometry.  
All measurements were made through
a red (approximately Johnson $R$) filter, 
with the exception of some images on 29 Aug that
were taken in both $V$ and $B$ to estimate stellar
colors.

The times at which a potential transit could occur were calculated
from the preliminary orbital period and ephemeris from M00.
The important elements were the orbital period $P$ 
and the time of maximum radial velocity of the star $T_{\rm max}$.
For this Letter, we have analyzed four nights of data; two of these
(29 Aug \& 13 Sep) occur off transit and establish the 
non-variability of the star,
while two (9 and 16 Sep) encompass the time of transit.
We produced calibrated images by subtracting a master bias and
dividing by a master flat.
Sixteen images from
16 Sep were averaged to produce a master image.  
We used DAOPHOT II (Stetson 1994)
to produce a master star list from this image, retaining
the 823 brightest stars.  
For each time series image, we then estimated a coordinate 
transformation, which allowed
for a linear shift $\delta x$ and $\delta y$.
We then applied
this coordinate transformation to the master star list
and carried out aperture photometry for all the images.  For each
star, a standard magnitude was defined from the result of the aperture
photometry on the master image.  
We corrected for atmospheric extinction using a color-dependent
extinction estimate derived from the magnitudes of
the 20 brightest stars in the field (excluding HD 209458 and two
obviously variable stars).
For two of the nights of
data (29 Aug \& 13 Sep), the residuals for HD 209458 are consistent
with no variation.  However, on the other
nights (9 Sep \& 16 Sep), we can see a conspicuous dimming of
the star for a time of several hours.  These residuals are
shown in Figure 1.  The root mean square (RMS) variation
in the resulting time series at the
beginning of the night of 9 Sep is 4 mmag;
the dominant source of noise for these bright stars is atmospheric
scintillation.  

\section{ANALYSIS OF LIGHT CURVE}
\subsection{Orbital Parameters}
As presented in M00, the derived orbital parameters from the combined 
radial velocity observations are $P = 3.52447 \pm 0.00029$ d and 
$T_{\rm max}=2451370.048 \pm 0.014$ HJD.

Since we observed two transits, it is possible to estimate independently both
a period and the time at the center of the transit, $T_c$, for the orbit.
To derive
the period, we phased the data to an assumed $P$ value,
in a range surrounding 3.5 d, 
and interpolated the data from the first transit
onto the grid of observation times for the later
transit.  The weighted sum of the square 
of the difference was calculated as a function
of assumed period, resulting in a clear minimum and
a well defined error.  We find the orbital period to be
$P = 3.5250 \pm 0.003$ d, consistent with but less precise than the
value determined from the radial velocity observations.  As
discussed in M00, the best fit
value of the mass for this star is $M_{s} = 1.1 M_{\sun}$; 
assuming this value, we
determine the semimajor axis to be $a = 0.0467$ A.U..

We used the data from the earlier transit,
which was the more precisely observed, to determine $T_c$.
For each assumed value of $T_c$,
we folded the light curve about $T_c$ and calculated the 
weighted sum of the square 
of the difference between the two halves of the folded curve.
We find that 
$T_{c} = 2451430.8227 \pm 0.003$ HJD. This value is consistent with
but is much more tightly constrained than the value 
determined from the radial velocity observations.  

Projecting the
errors in $P$ from the radial velocity observations and 
$T_{c}$ from the photometry observations, the time of transit
can be calculated with a precision of better than half an hour
for the next six months.

\subsection{Interpretation of the Transit Curve}
For the purpose of interpreting the light curve,
we binned the residuals from both transits into 
5 minute time bins according to
the orbit derived above.  The time series RMS 
of these binned data is 1.5 mmag throughout the
timespan covered by the observations, with an increase
to larger scatter roughly 1 hour after the point of last contact
due to the increasing airmass.  These binned data are plotted in Figure 2.

Five parameters participate in determining
the precise shape of the transit curve.  These are the planetary
radius $R_{p}$, the stellar radius $R_{s}$, the stellar
mass $M_{s}$, the orbital
inclination angle $i$, and the limb darkening parameter $c_{\lambda}$,
where the normalized stellar surface brightness profile is written as 
$B_{\lambda}(\mu)=1-c_{\lambda}(1-\mu)$, and $\mu$ is 
the cosine of the angle between the normal to the stellar surface
and the line-of-sight.  Though they are physically distinct,
not all of these parameters have independent influences on the light curve.
To estimate these parameters, we adopt the following approach.
Absent photometric observations, there are no observational data
to restrict $R_p$ and $i$, whereas there is a wealth of
information on $R_s$, $M_s$, and $c_{\lambda}$ from the 
theory of stars and the observed
values of the star's brightness, color, metallicity, temperature,
and age.  A detailed presentation of the best fit values
for $R_s$, $M_s$, and $c_{\lambda}$ is given in M00.
From a preliminary investigation, 
we adopt $R_{s} = 1.1 \ R_{\sun}$ and $M_{s} = 1.1 \ M_{\sun}$,
consistent with the G0 V spectral type of the star.
The photometry we present here was taken in the $R$ band, for
which we take $c_{R}=0.5$, the solar value (Allen 1973).

We modeled the data as follows.  For each
assumed value of \{$R_{p}$,$i$\}, we calculated the relative flux change 
at the phase of each observation by 
integrating the flux occulted by a planet
of given radius at the correct projected location
on a limb-darkened disk, computing the
integral over the unobstructed disk, and forming the ratio.
A detailed description of the
form of this integral can be found in Sackett (1999).
We then calculated the ${\chi}^2$ of the model applied to the
5 m phase binned time series.  A contour plot
of this $\chi^2$ surface appears in Figure 3.  

Adopting \{$R_{s}$,$M_{s}$\} = \{1.1, 1.1\} (solar units),
the best fit values for the parameters are 
$R_{p}=1.27 \pm 0.02 \  R_{\rm Jup}$ and 
$i = 87.1 \pm 0.2 ^{\circ}$.  The uncertainties quoted correspond to
the 1-$\sigma$ contour in the ${\chi}^2$ surface.
These are not, however, correct estimates of the true uncertainties. 
The uncertainty in the physical parameters
of the star, namely $R_{s}$, $M_{s}$, and ${c_{R}}$, 
cause uncertainties in $R_p$ and in $i$
that exceed the stated formal errors.  The systematic effect
due to an uncertainty in either $M_{s}$ or ${c_{R}}$ is small.
In particular, holding the stellar radius and mass constant, and 
changing the value of ${c_{R}}$ by $\pm 0.1$ 
changes the best-fit value of $i$ by $\pm 0.15^{\circ}$ and 
$R_{p}$ by $\pm 0.01 \ R_{\rm Jup}$.  In contrast to this, 
the effect due to the uncertainty in $R_{s}$ is significant.
In Figure 3, we show the confidence ellipses for 
several choices of stellar mass and radius.  
This demonstrates the fashion in which a larger
star would require a larger planet at a lower inclination
to adequately fit the data.
We note, however, that no plausible values of the stellar radius or mass
give $R_{p}$ as large as $1.6 \ R_{\rm Jup}$, the value presented by
Henry et al. (1999).

\section{DISCUSSION}
Since we have measured the orbital inclination, 
we can estimate the true mass of the planet.
Using our measured value $i = 87.1^{\circ}$,
we derive $M_{p} = 0.63 \ M_{\rm Jup}$.

The derived value of $R_{p}=1.27 \ R_{\rm Jup}$ is in excellent
agreement with the early predictions of Guillot et al. (1996),
who calculated the radius for a strongly irradiated 
radiative/convective extrasolar planet for a variety of masses.

Since this is the first extrasolar planet of a known
radius and mass, several physically important
quantities can be calculated for the first time.
We estimate an average
density of $\rho \approx 0.38 \ {\rm g \, cm^{-3}}$, 
which is significantly less than the density of Saturn,
the least dense of the solar system gas giants.
The surface gravity
is $g \approx 970 \ {\rm cm \, s^{-2}}$.  
Assuming an effective temperature for the star of 
$T_{s}=6000 K$, and a planetary albedo of $A$, 
the effective temperature of the planet
is $T_{p} \approx 1400 \ (1-A)^{\frac{1}{4}} \ K$.  
This implies a thermal velocity
for hydrogen of $v_{t} \lesssim 6.0 \ {\rm km \, s^{-1}}$.  
This is roughly a factor of 7
less than the calculated escape velocity of $v_{e} \approx 
42 \ {\rm km \, s^{-1}}$,
confirming that these planets should not be losing
significant amounts of mass due to the effects of 
stellar insolation.

The existence of a transiting planet suggests many fruitful
observations that bear on both planetary and stellar physics.

It will be highly desirable to obtain high cadence photometry
of similar precision in as many band passes as possible.
Observing the color dependence of the transit shape
will measure the limb darkening, and break the degeneracy
shared between this effect and the other parameters.  In particular,
we predict a deeper transit in $V$ and $B$, due to greater
limb darkening at these shorter wavelengths.

If there are other planets in the HD 209458 system, and if their orbits
are approximately coplanar with the one we observe, then the likelihood that
they too will generate transits is substantially enhanced relative to
that for a randomly oriented system.
The radial velocity data of M00 do not suggest other massive objects in this
system.
However, less massive objects (similar to Uranus, for
instance) could easily escape detection via radial velocities,
and yet be observable photometrically.
Gilliland and Brown (1992) showed that observations with a 2-m telescope could
attain precision of 400 $\mu$mag per minute of integration.
A central transit by a Uranus-sized planet at 0.2 AU would yield
a dimming some 6 hours in duration, with a depth of about 1 mmag;
this would be easily visible with the abovementioned precision.
With the accuracy that should be attainable from outside the Earth's
atmosphere (eg., Borucki et al. 1997), planets the size of Earth
would be detectable.
Note, however, that in the case of exactly coplanar orbits, the orbital
inclination of 87.1$^\circ$ implies that planets further than 0.1 AU
from the star will show no transits.
From this point of view, a small dispersion in orbital inclination would
enhance, not decrease, the chances of observing other planets.

Other objects in the HD209458 system need not be separate planets;
they could be moons of the known planet, or even dust rings surrounding
it.
Dynamical considerations restrict the allowable distance of such bound
objects from the planet, simplifying the detection problem.
Precise photometric searches for such objects could in principle yield
detections for moons only slightly larger than the Earth.

Reflected light observations such as those for the $\tau$ Boo system 
by Charbonneau et al. (1999) and Cameron et al. (1999) 
will be difficult, because of the relative
faintness of this star.
If successful, however, they would yield the planet's albedo directly,
since its radius is accurately known.
In particular, the ratio of the flux from the planet at opposition to 
that of the star is $\sim 1.7 \times 10^{-4} \ p_{\lambda}$,
where $p_{\lambda}$ is the wavelength dependent geometric albedo. 
Similarly, observations at wavelengths longer than a few microns may
detect the secondary eclipse as the planet passes behind the star.
In the long wavelength limit, the depth ratio of the primary to the secondary
eclipse should be the ratio of the effective temperatures, roughly 4,
leading to signals of perhaps 3 mmag.
Knowing this depth would allow the planet's actual dayside temperature
to be estimated, hence constraining the mean atmospheric absorptivity.

Differences between spectra of the star in and out of transit would
reveal changes in the line depths at roughly the 1\% level due
to the variations in the line profile integrated over the
visible surface of the partially occulted star.  At a level
of roughly 0.01\%, it may be possible to observe absorption features
added by the planetary atmosphere while it is in transit.
The atmospheric scale height will be greatly enhanced due
to the high temperature of the planet, and this effect
may bring the amplitude of the absorption features into
the observable regime.  Charbonneau et
al. (1999) have demonstrated that time varying changes in the
spectrum of the system can be monitored at the level of better
than $5 \times 10^{-5}$ for the bright star $\tau$ Boo.  If this technique 
can be extended to this star, it would be possible to search for these
absorption features.
 
\section{CONCLUSION}
The discovery of transits in the light curve of HD209458 confirms
beyond doubt that its radial velocity variations arise from an orbiting
planet.
Moreover, having a reliable estimate of the planetary radius and mass,
we can now say with assurance that the planet is indeed a gas giant.
We are encouraged by the closeness of the fit between our observed
radius and that computed on theoretical grounds by Guillot et al. (1996).
One may nonetheless expect that observations of this and similar systems
will rapidly become more sophisticated and penetrating, and that the
results will be puzzling more often than not.
We are confident that solving these puzzles will lead to a new and far
more comprehensive understanding of the processes that control the formation
and evolution of planets,
and to an exciting time for those with an interest in the field.

\acknowledgements
The identification of HD 209458 as a prime target for transit observations
was made possible by the many contributions of the members of the G Dwarf
Planet Search, ELODIE, and CORALIE teams:  J. L. Beuzit, M. Burnet, G. A.
Drukier, T. Mazeh, D. Naef, F. Pepe, Ch. Perrier, D. Queloz, N. Santos,
J. P. Sivan, G. Torres, S. Udry, and S Zucker.  We are especially grateful
to T. Mazeh and the members of his team in Tel Aviv for their critical
role in the analysis of the Keck observations.
We are grateful to R. Noyes for many helpful conversations, 
and we thank the referee E. Dunham for helpful comments, 
which improved the paper.  
Furthermore, we thank G. Card, C. Chambellan, 
D. Kolinski, A. Lecinski, R. Lull,
T. Russ, and K. Streander
for their assistance in the fabrication, maintenance, and operation of
the STARE photometric camera.  We also thank P. Stetson for the
use of his DAOPHOT II software.  
D. Charbonneau is supported in part by a Newkirk Fellowship of the High
Altitude Observatory.
This work was supported in part by NASA grant W-19560.


\begin{figure}
\plotone{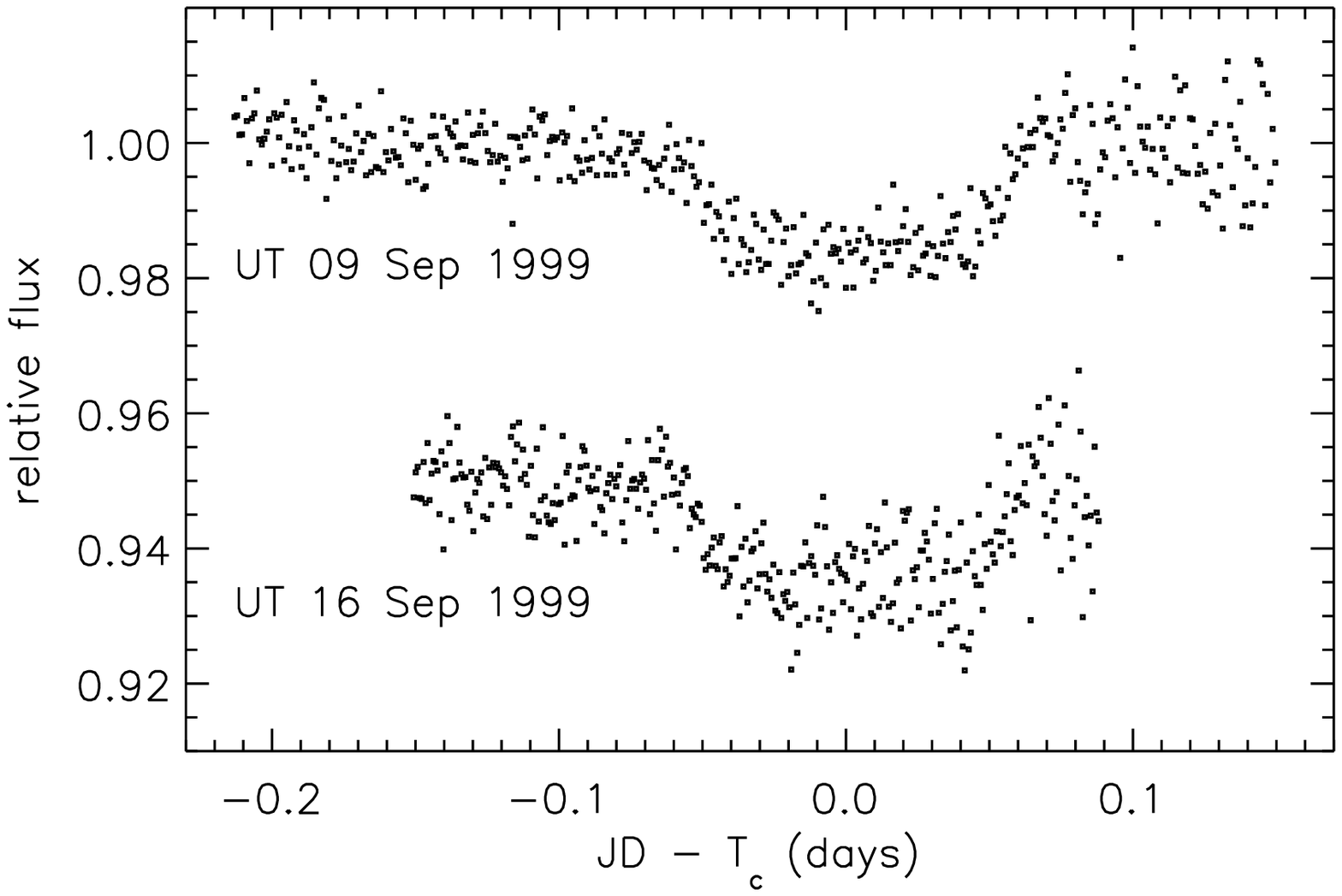}
\figcaption{Shown are the photometric time series, corrected for 
gray and color-dependent extinction, for 9 \& 16 Sep 1999,
plotted as a function of time from $T_{c}$.  The
RMS of the time series at the beginning of the night
on 9 Sep is roughly 4 mmag.
The increased scatter in the 16 Sep data relative to the
9 Sep data is due to the shorter exposure times.
The data from 16 Sep are offset by -.05 relative to those from 9 Sep.}
\end{figure}

\begin{figure}
\plotone{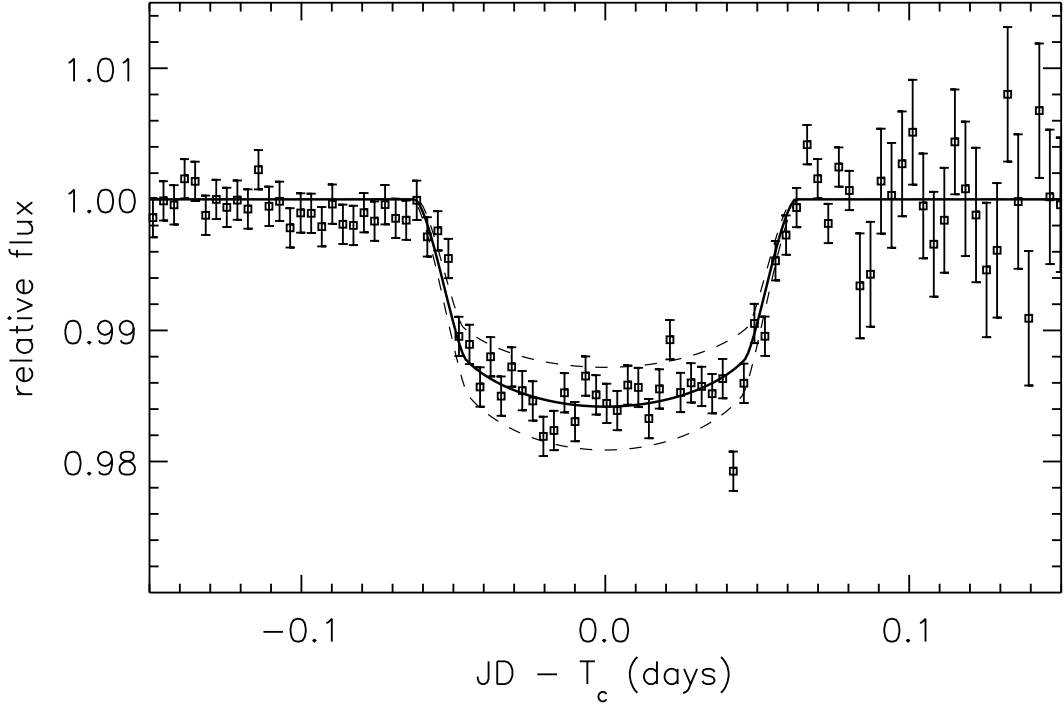}
\figcaption{Shown are the data from figure 1 binned into 5 m averages,
phased according to our best-fit orbit,
plotted as a function of time from $T_{c}$.
The RMS variation at the beginning of the time series
is roughly 1.5 mmag, and this precision
is maintained throughout the duration of the transit.  The increased
scatter at the end of the timeseries is due to increasing airmass 
which occured at roughly the same time for both transits,
since the two
occured very nearly one week apart.  The solid line is the transit
shape that would occur for our best fit model, $R_{p} = 1.27 \ R_{\rm Jup}$,
$i = 87.1^{\circ}$.  The lower and upper dashed
lines are the transit curves that would occur for a planet 
10\% larger and smaller in radius, respectively.  The rapid initial fall 
and final rise of the transit curve correspond
to the times between first and second, 
and between third and fourth contacts, when
the planet is crossing the edge of the star; the resulting slope is a function
of the finite size of the planet,
the impact parameter of the transit, and the limb darkening of the star.
The central curved portion of the transit is the time between 
second and third contacts, when the planet is entirely in front of
the star.}
\end{figure}

\begin{figure}
\plotone{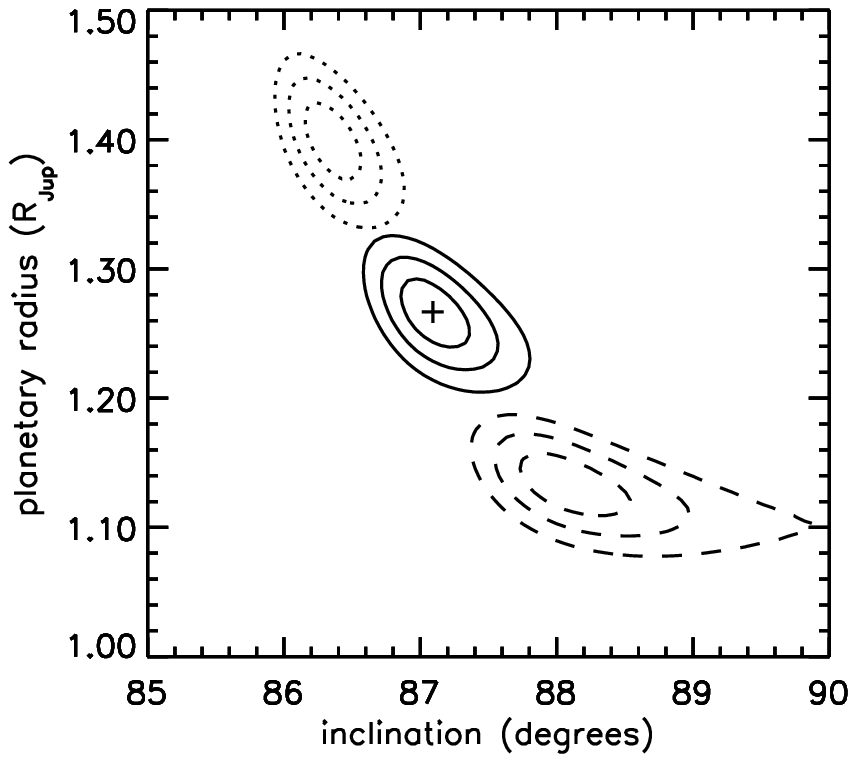}
\figcaption{The solid contours are the 1-, 2-, \& 3-$\sigma$
confidence levels for the planet radius and orbital inclination,
assuming $R_{s} = 1.1 \ R_{\sun}$ and
$M_{s} = 1.1 \ M_{\sun}$.  The minimum occurs at 
$R_{p} = 1.27 \ R_{\rm Jup}$ and $i = 87.1^{\circ}$.
The dashed and dotted contours are the confidence levels
in the cases of \{$R_{s},M_{s}$\} $=$ \{1.0 $R_{\sun}$,1.0 
$M_{\sun}$\} and \{$R_{s},M_{s}$\} $=$ \{1.2 $R_{\sun}$,1.2 
$M_{\sun}$\} respectively.  The dominant modeling uncertainty 
is that in the stellar radius.}
\end{figure}


\begin{references}
\reference{p1}Allen, C. W. 1973, Astrophysical Quantities, 3rd ed,
Athlone Press, London
\reference{pn}Baranne, A., Queloz, D., Mayor, M., Adrianzyk, G., 
Knispel, G., Kohler, D., Lacroix, D., Meunier, J.-P., Rimbaud, G., 
\& Vin, A. 1996, A\&AS, 119, 373
\reference{p2}Borucki, W., Koch, D., \& Webster, L. 1997,
Kepler Mission: A Search for Habitable Planets, proposal to NASA
\reference{p2i}Brown, T. M. \& Kolinski, D. 1999, http://www.hao.ucar.edu/public/research/stare/stare.html
\reference{p3}Butler, R. P., Marcy, G. W., Williams, E., Hauser, H.,
\& Shirts, P. 1997, \apjl, 474, L115
\reference{p4}Butler, R. P., Marcy, G. W., Vogt, S. S., 
\& Apps, K. 1998, \pasp, 110, 1389
\reference{p5}Cameron, A., Horne, K., Penny, A., \& James, D. 1999, \nat, in press
\reference{p6}Charbonneau, D., Noyes, R. W., Korzennik, S. G., Nisenson, P., 
Jha, S., Vogt, S. S., \& Kibrick, R. I. 1999, \apjl, 522, L145
\reference{p7}Fischer, D. A., Marcy, G. W., Bulter, R. P., Vogt, S. S., \& 
Apps, K. 1999, \pasp, 111, 50
\reference{p8}Gilliland, R.L \& Brown, T.M. 1992, \pasp, 104, 582
\reference{p9}Guillot, T., Burrows, A., Hubbard, W. B., Lunine, J. I., 
\& Saumon, D.  1996, \apjl, 459, L35
\reference{p10}Henry, G., Marcy, G., Butler, R. P., \& Vogt, S. S. 1999,
IAU Circ 7307 
\reference{p11}Mayor, M., Naef, D., Udry, S., Santos, N., Queloz, D., 
Melo, C., \& Confino, B. 1999, 
http://obswww.unige.ch/$\sim$udry/planet/hd75289\_ann.html
\reference{p12}Mayor, M., \& Queloz, D. 1995, \nat, 378, 355
\reference{p13}Mazeh, T. et al. 2000, \apjl, in preparation (M00)
\reference{p15}Queloz, D., et al. 1999, A\&A, accepted, 
preprint available at http://xxx.lanl.gov/abs/astro-ph/9910223
\reference{p14}Sackett, P. D. 1999, in NATO/ASI Ser., Planets Outside
the Solar System: Theory and Observations, ed. J.-M. Mariotti \& D. Alloin (Dordrecht: Kluwer), 189
\reference{p15}Stetson, P. B. 1994, \pasp, 106, 250
\reference{p16}Udry, S., Mayor, M., Naef, D., Pepe, F., 
Queloz, D., Santos, N., \& 
Burnet, M. 1999, http://obswww.unige.ch/$\sim$udry/planet/hd130322\_ann.html
\reference{p20}Vogt, S. S. et al. 1994, Proc. SPIE, 2198, 362
\end{references}
\end{document}